\documentclass[sigconf]{acmart}

\usepackage{graphicx}
\usepackage{multirow}
\usepackage{array}
\usepackage{subcaption}
\usepackage{caption}
\AtBeginDocument{%
  }
    
\copyrightyear{2024}
\acmYear{2024}
\setcopyright{acmlicensed}\acmConference[ISLPED '24]{Proceedings of the ACM/IEEE International Symposium on Low Power Electronics and Design}{August 5--7, 2024}{Newport Beach, CA, USA}
\acmBooktitle{Proceedings of the ACM/IEEE International Symposium on Low Power Electronics and Design (ISLPED '24), August 5--7, 2024, Newport Beach, CA, USA}
\acmDOI{10.1145/3665314.3670816}
\acmISBN{979-8-4007-0688-2/24/08}

\begin{document}
\title{Hyft: A Reconfigurable Softmax Accelerator with Hybrid Numeric Format for both Training and Inference}

\author{Tianhua Xia}
\affiliation{%
  \institution{New York University}
  \city{New York}
  \country{USA}}
\email{tx856@nyu.edu}

\author{Sai Qian Zhang}
\affiliation{%
  \institution{New York University}
  \city{New York}
  \country{USA}}
\email{sai.zhang@nyu.edu}




\begin{abstract}
The attention mechanism is a pivotal element within the transformer architecture, making a substantial contribution to its exceptional performance. Within this attention mechanism, Softmax is an imperative component that enables the model to assess the degree of correlation between various segments of the input. Yet, prior research has shown that Softmax operations can significantly increase processing latency and energy consumption in the transformer network due to their internal nonlinear operations and data dependencies.
In this work, we proposed~\textit{Hyft}, a hardware efficient floating point Softmax accelerator for both training and inference. Hyft aims to reduce the implementation cost of different nonlinear arithmetic operations within softmax by adaptively converting intermediate results into the most suitable numeric format for each specific operation, leading to reconfigurable accelerator with hybrid numeric format. The evaluation results highlight that Hyft achieves a remarkable $10\times$ reduction in hardware resource utilization and a $6 \times$ reduction in processing latency, all while maintaining a negligible impact on transformer accuracy.


\end{abstract}


  
\maketitle
\section{Introduction}
Ever since the debut of the transformer~\cite{vaswani2017attention}, attention-based deep neural networks (DNNs) have achieved remarkable performance across various tasks in different fields~\cite{devlin2019BERT, dosovitskiy2021image}. The attention mechanism in the transformer plays a crucial role in enabling the model to capture and model complex relationships and dependencies within input sequences, greatly enhancing the accuracy on different tasks. 
\begin{figure}
\captionsetup{font=small}
    \centering
    \hspace*{-1\baselineskip}
    \includegraphics[width=0.5\textwidth]{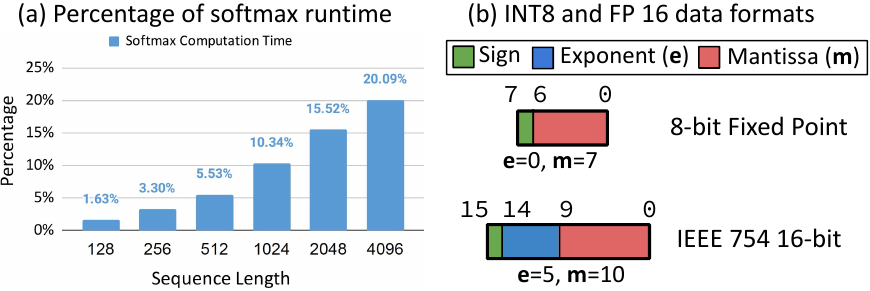}
    \caption{(a) Percentage of Softmax computation runtime in BigBird. (b) Fixed and floating-point number formats.}
    \label{fig:softmax_runtine}
\vspace*{-1\baselineskip}
\end{figure}

Nonetheless, the superior accuracy of transformer is accompanied by increased computational and memory requirements. The widespread development of large language models (LLMs) has led to transformer sizes reaching unprecedented scales and continuing to expand. On top of the model size, attention computation consists of a unique mix of linear matrix multiplication and non-linear operations such as Softmax, layer normalization, and GeLU. Unlike other DNN architectures like Convolutional Neural Networks (CNNs) and Recurrent Neural Networks (RNNs), transformers allocate a substantial portion of their runtime to attention operations.
In particular, recent work has shown that Softmax operations can consume a large portion of runtime in the transformer, especially as the length of input sequences increases (>30\%~\cite{softermax}). This impact is not limited to transformer model for NLP tasks. It also affects variants like vision transformers (ViT)~\cite{dosovitskiy2021image}, and large models~\cite{openai2023gpt4}. 

To demonstrate the cost of Softmax in transformer model, we execute the Huggingface BigBird model \cite{wolf2020huggingfaces} with half-precision (FP16) on an A100 GPU, and profile the Softmax GPU computation time under different input sequence lengths.
As shown in Figure~\ref{fig:softmax_runtine} (a), Softmax computation takes up a large amount of model run time, especially when processing longer sequences ($\approx 20\%$ for a input length of 4096). 


Previous research has revealed a notable contrast: unlike other DNN models (e.g., CNNs) that exhibit insensitivity to Softmax computation, the performance of transformers is notably influenced by the Softmax output~\cite{klut, isca_sm}. This sensitivity implies that many current approximation algorithms for Softmax, such as those that reduce the precision of Softmax operands~\cite{zfw2018,aggresive_sm}, can result in a serious accuracy degradation on the transformers. 

Accelerating the Softmax operation presents a significant challenge. Specifically, the complexity of hardware implementation for Softmax arises from two main factors. Firstly, the exponential and division operations involved in Softmax are computationally intensive and resource-demanding. Secondly, the data dependency in Softmax makes it difficult to pipeline and parallelize the operations effectively, resulting in increased latency and energy consumption. 
Furthermore, to maintain compatibility with the other computations within attention-based models, it is preferable for the Softmax block to operate with floating-point input and output formats.
In addition to inference operations for transformers, there is an increasing demand for fine-tuning Transformer models, particularly transformer-based Large Language Models (LLMs), to dynamically adapt to unseen user datasets and tasks~\cite{LoRA,AdapterFusion}. This underscores the significance of effectively conducting Softmax training.


In this study, we introduce~\textit{Hyft}, a hardware-efficient Softmax accelerator designed for both training and inference. Hyft is designed to reduce the implementation cost of various nonlinear arithmetic operations by dynamically converting intermediate results into the most appropriate numeric format for each specific operation. Additionally, Hyft offers a high degree of reconfigurability, allowing users to adjust input and output precision to achieve an optimal balance between hardware performance and accuracy. Moreover, to facilitate the backward propagation operations for Softmax, Hyft strategically reuses the hardware blocks for forward propagation. This effectively increases the hardware utilization rate, thereby enhancing the efficiency of training operations. Finally, Hyft is a plug and play device that is compatible with other transformer acceleration techniques (e.g., SpAtten~\cite{wang2021spatten}, FlashAttention~\cite{dao2022flashattention}).  To summarize, our contributions can be outlined as follows:

\begin{itemize} 
\item To the best of our knowledge, Hyft marks the first Softmax accelerator capable of efficiently supporting both Softmax inference and training operations in transformers.
\item Hyft leverages a hybrid approach, combining fixed and floating-point formats, to streamline the exponential, division, and multiplication operations involved in both the forward and backward propagation of Softmax.
\item Hyft offers a high degree of configurability, allowing users to choose the optimal numeric format for representing intermediate results. This flexibility enables a superior trade-off between hardware performance and accuracy performance.
\item The evaluation results highlight that Hyft achieves a remarkable $10\times$ reduction in hardware resource utilization and a $6 \times$ reduction in processing latency, all while maintaining a negligible impact on transformer accuracy.

\end{itemize}



\section {Background and Motivation}
\subsection{Numeric Format}

Numerous numeric formats have been investigated to alleviate the computational demands in DNN inference and training, as documented in previous studies~\cite{mcdanel2019full, zhang2022fast, zhang2023camel}. These formats can broadly be classified into two main categories: fixed-point formats and floating-point formats. Fixed-point formats lack an exponent field, which results in a reduced dynamic range representation but simplifies the hardware implementation. The upper part of Figure~\ref{fig:softmax_runtine} (b) show an example on fixed-point representation with 8 bits.

On the contrary, floating-point formats offer a wider dynamic range when compared to fixed-point formats. While the conventional format for DNN training is IEEE 754 32-bit floating point or FP32, there has been exploration into other options such as 16-bit floating point (FP16) and various custom floating-point formats like bfloat16 and TensorFloat. The lower part of Figure~\ref{fig:softmax_runtine} (b) displays the IEEE 754 half-precision floating-point format (FP16). The value of a floating-point number can be calculated using Equation \ref{ieee754_val_cal}:
\begin{equation} \label{ieee754_val_cal}
    x = (-1)^{S_x}2^{e_x}(1+m_x) 
\end{equation}
where $S_x$ represents the sign bit, $e_x=E_x-B$ denotes the value of the 5-bit exponent as $B$ represents the bias, and $m_x = M_x/2^{L}$ stands for the value of the mantissa bits.
In the IEEE 754 half-precision format, $B$ is set to 15, and $L$ is set to 10. 

\subsection{Computation in Transformer}

A transformer is constructed as a stack of transformer encoders and decoders, with each encoder or decoder comprising two fundamental components: a Self-Attention (SA) block and a Multi-Layer Perceptron (MLP) block. During the inference process, the input $H$ of the SA block is first multiplied with three weight matrices $W_{Q}$, $W_{K}$, and $W_{V}$, yielding the outputs referred to as query ($A_Q$), key ($A_K$), and value ($A_V$), respectively. 
The resulting $A_{Q}$ and $A_{K}$, in combination with $A_{V}$, will then undergo multiplication, Softmax, and residual addition to generate the SA output $H_{out}$ in equation~\ref{subeq:query2}, where $d_{head}$ indicate the number of feature dimensions in multi-head attention mechanism~\cite{vaswani2017attention}.
\begin{equation}
    H_{out} = W_o \times (A_v \times Softmax(\frac{A_Q \times A_K^T}{\sqrt{d_{head}}})) \label{subeq:query2}
\end{equation}

The SA output will then be forwarded to the MLP blocks for further processing. The feed forward block consists of a stack of fully connected (FC) layers together with some intermediate activation function. The output of the last decoder layer will be sent to a linear layer, which then generates the final results. 

\subsection{Softmax Function}
Softmax operation has been widely utilized in the transformer architecture. Specifically, the Softmax function takes a vector $\boldsymbol{z}=[z_1, z_2,\cdots,z_N]^T$ and generates an output $\boldsymbol{s}=[s_1, s_2,\cdots,s_N]^T$, both have a length of N. It is defined as follows:
\begin{equation}
s_i=\frac{e^{z_i}}{\Sigma_{j=0}^{N-1}e^{z_j}} ~~ 
For~i=1,2,\cdots,N
\end{equation}
direct calculations may encounter numeric issues due to limitations of the 32-bit single-precision floating-point format further leading to a NaN (Not a Number) outputs.
To ensure numeric stability, the Softmax function's inputs are typically calibrated by subtracting their max from each input, as illustrated below:
\begin{equation} 
\label{Softmax_2}
\begin{split}
s_i=\frac{e^{z_i-z_{max}}}{\Sigma_{j=0}^{N-1}e^{z_j-z_{max}}} ~~ 
For~i=1,2,\cdots,N
\end{split}
\end{equation}
This operation results in a stable input with a consistent range, without altering the final results. Additionally, the back propagation computation of Softmax is defined as follows:


\begin{equation} 
\label{sm_derivative}
\frac{\mathrm{d}\boldsymbol{s}}{\mathrm{d} \boldsymbol{z}} = diag(\boldsymbol{s})-\boldsymbol{s}\boldsymbol{s}^T
\end{equation}
where $diag(s)$ represents a square matrix with $s$ as its diagonal elements, while all other elements are set to zero. An example is shown below when $N=3$:

\[
\small
\frac{\mathrm{d}\boldsymbol{s}}{\mathrm{d} \boldsymbol{z}}
=
 \begin{bmatrix} 
s_1 -s_1^2 & -s_1 \cdot s_2 & -s_1 \cdot s_3\\
-s_2 \cdot s_1 & s_2-s_2^2 & -s_2 \cdot s_3\\
-s_3 \cdot s_1 & -s_3 \cdot s_2 & s_3 -s_3^2
\end{bmatrix}
\]    

\subsection{Related Work}

Numerous approximation algorithms have been proposed for the Softmax operation for the last layer of CNN, such as Taylor expansion~\cite{nuaa} and division approximation~\cite{zfw2018}.
Additionally, in works like \cite{softermax} and \cite{base2sm}, authors have simplified Softmax implementation by replacing the base-e Softmax function with a base-2 version. However, it is important to note that all these previous solutions necessitate fine-tuning the DNN to account for the approximation error. Furthermore, all of these approaches employ fixed-point or low-precision floating-point formats for value storage, which deviates from the numeric format used in the remaining layers of the network. 
Additionally, it is important to note that none of the aforementioned methods support backpropagation operations for the Softmax layers in transformer. While~\cite{base2sm} does support training operations, it necessitates fine-tuning the resulting CNN to mitigate the significant approximation error introduced by the base-2 Softmax approximation. 

In contrast, Hyft operates using a half-precision mode to accommodate FP16 input and output, and it can be readily configured to function in full-precision mode, supporting input and output in FP32. Hyft can be seamlessly integrated into any existing DNNs without the need for format conversion.

\section{Hyft Architecture}
\label{sec:hyft-arch}
\begin{figure}
\captionsetup{font=small, justification=centering}
\hspace*{-1\baselineskip}
    \centering   \includegraphics[width=0.45\textwidth]{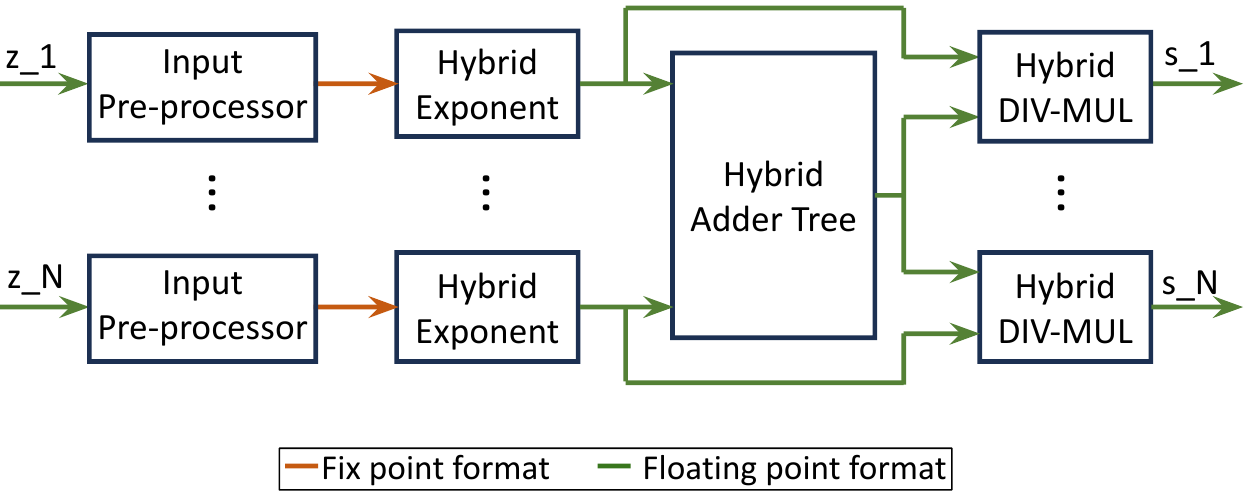}
    \caption{Forward propagation data path of Hyft.}
    \label{fig:sm_arch}
 \vspace*{-1\baselineskip}
\end{figure}

The overall computation flow of Hyft during the forward propagation is illustrated in Figure~\ref{fig:sm_arch}. The Hyft system comprises four core components: input pre-processor, hybrid exponent unit,  hybrid adder tree, and hybrid multiplication/division unit.

Hyft supports input and output data in either FP16 or FP32 and dynamically converts intermediate data to the most suitable number format for the current arithmetic operation. In Figure~\ref{fig:sm_arch}, you can observe the floating-point data paths highlighted in green and the fixed-point data paths highlighted in red.
To achieve this, we utilize a fixed-point format for internal steps to streamline linear addition and subtraction operations. Additionally, we employ floating-point representation to facilitate exponential, multiplication, and division operations within the logarithmic space. The precision of the intermediate results can be fully configurable. 



Given the input in floating-point format $\boldsymbol{z}=[z_1, z_2,\cdots,z_N]^T$, the input pre-processor operates by identifying the maximum value, denoted as $z_{max}$, within the input vector. To simplify the later input subtraction operation, the each element $z_{i}$ of $\boldsymbol{z}$ and $z_{max}$ will be converted to fixed-point format with configurable precision in the pre-processor. The hybrid exponent unit subtracts $z_{max}$ from each element in the input vector, producing  $\boldsymbol{z'}=[z'_1, z'_2,\cdots,z'_N]^T$, where $z'_i = z_{i}-z_{max}$. $z'_{i}$ expressed in fixed point format will benefit the hybrid exponent unit by simplifying the integer fraction split step, which will be described in Section~\ref{sec:hybrid_exp_section}. The Hybrid exponent unit processes input data in fixed-point format and produces output data in floating-point format. This approach eliminates the need for shift operations, which is typically needed in traditional fixed-point exponential accelerators. The hybrid adder tree converts the floating-point inputs to fixed-point format to simplify the addition operations. After performing the additions, it then converts the results back to floating-point format. This approach helps streamline the addition process while maintaining compatibility with floating-point representations. The division unit performs floating point division and generates the Softmax results in floating point.

During our FPGA implementation, we observe that without the numeric format conversion among the blocks, the shift operations in the hybrid exponent units will become the critical path for Softmax computation. By strategically employing hybrid formats at various stages, we can significantly reduce the number of shift operations required, thus optimizing the overall performance. Next, we will introduce each component in detail.

\subsection{Parameterized Input Pre-Processr} \label{sec:input_norm_src}
\begin{figure}
\captionsetup{font=small}
    \centering
    \includegraphics[width=0.48\textwidth]{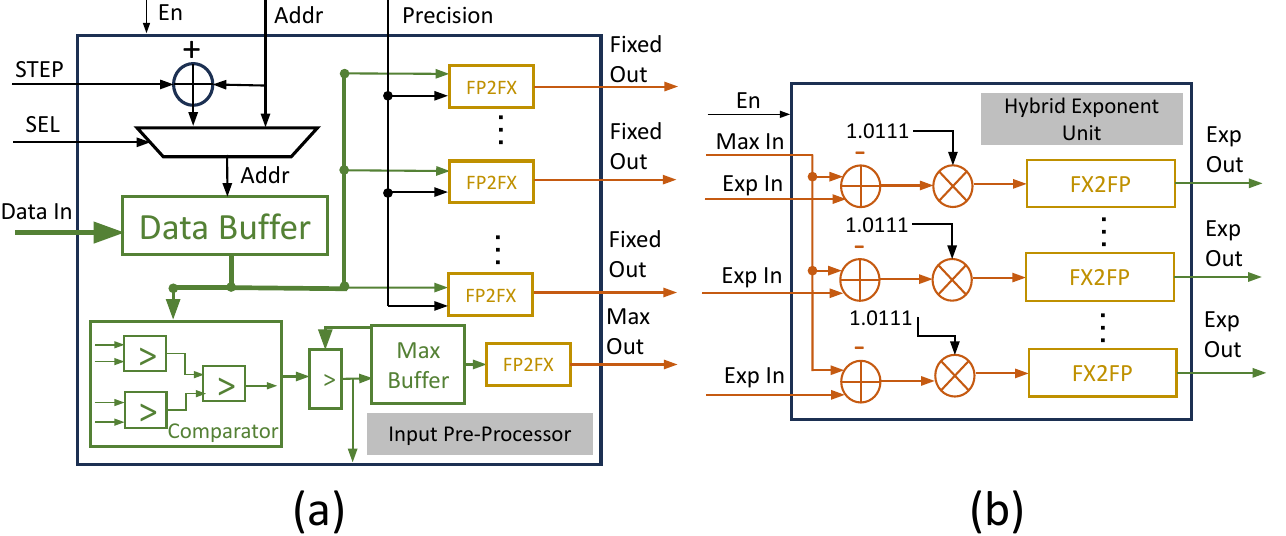}
    \caption{(a) Input Pre-processor and (b) Hybrid Exponent Unit. The floating-point and fixed-point data paths are highlighted in green and red, respectively. The control signal are highlighted in black.}
    \label{fig:in_norm}
\vspace*{-1\baselineskip}
\end{figure}

The parameterized input pre-processor serves two essential functions. 
Firstly, it searches the maximum value of the input vector to be used in hybrid exponential unit to prevent Softmax calculation overflow issues.
Secondly, it converts floating-point inputs to fixed-point format to align with the input format of hybrid exponent unit (Section~\ref{sec:hybrid_exp_section}), simplifying hardware implementation as discussed later. The architecture of the input pre-processor is depicted in Figure~\ref{fig:in_norm} (a).

To accelerate the search for the maximum value among $z_{i}$, we leverage on the resilience of the transformer architecture to computational noise. Specifically, a configurable parameter,~\textit{STEP}, is taken by input pre-processor as a input, which governs the number of data points that enter the comparator block for the max value search. When set to 1, the address of the next data to be compared is the current address incremented by 1, causing Hyft to utilize all the data for the max search. In contrast, when ~\textit{STEP} is set to 2, the address of the next data is the current address incremented by 2, leading Hyft to use every other data point for the max search. We show in the evaluation section that most of the tasks can be executed with an accelerated maximum searching process without any accuracy degradation. 

In parallel to max search block, there are floating point to fixed point converters (FP2FX) in the input pre-processor which converts the floating point inputs and their max value to fixed point format. The fixed point format outputs will benefit the hybrid exponent unit in Section~\ref{sec:hybrid_exp_section}. The input preprocessor in Hyft also accepts an additional parameter, denoted as ~\textit{Precision}, which determines the number of bits allocated for the decimal part in the converted fixed-point format.

\subsection{Hybrid Exponent Unit} \label{sec:hybrid_exp_section}

Our hybrid exponent unit is designed to accept fixed-point numbers $z$ and $z_{max}$ as input and produce results $e^{z-z_{max}}$ in floating-point format. Utilizing fixed-point input simplifies the input subtraction, the constant multiplication and the separation of integer and decimal components within our exponential unit. The use of floating-point output eliminates the delays associated with shift operations found in traditional fixed-point exponential accelerators. Importantly, this design also enhances the performance of the division and multiplication units within our Softmax implementation.

Exponent operations $e^{z'}$ involve numerous multiplications that consume large computational resources and lead to increased processing latency. To eliminate the computations associated with the Euler number, we transform the exponential operation as follows:  
\begin{equation} 
\label{exp2shift}
e^{z'}=2^{z'log_2e}=2^{u+v}
\end{equation}
where $u$ and $v$ represent the integer and fractional parts of $z'log_{2}e$ value. We approximate $log_2e$ as $1.0111_{2}$, therefore $z'log_2e$ can be approximated as $z' +(z'>>2)+(z'>>3)+(z'>>4)$. We further apply Booth encoding, reducing the number of shift operations required, resulting in the approximation $z'log_2e\approx z' +(z'>>1)-(z'>>4)$. This shift-and-add operation is efficiently computed using an adder trees. Leveraging the inherent characteristics of the fixed-point format, it becomes straightforward to extract both the integer and decimal parts of $z' \cdot log_2e$. 

Due to the input normalization described in Equation \ref{Softmax_2}, the input to the exponential unit is consistently less than or equal to zero. Consequently, the values represented by $u,v$ in Equation \ref{exp2shift} are always less than or equal to zero as well. Building upon the insights presented in~\cite{isca_sm}, we can further streamline the computation using the Taylor approximation, which is shown as follows:
\begin{equation} 
\label{u_v}
e^{z'}=2^{u+v} \approx 2^u(1+v/2)   \enspace\enspace\enspace\enspace u \leq 0,-1 <v \leq 0
\end{equation}

To be precise, the current intermediate results $u$ and $v$ are less than or equal to zero, but floating point format requires mantissa being a positive number. A simple FX2FP block  is used to convert the exponent and mantissa fields for the floating-point representation $e^{z'}_{float}$ of $e^{z'}$ using the following formula:
\begin{equation} 
\label{uv}
e^{z'} \approx 2^u(1+v/2) = 2^{u-1}(1+(1+v))
\end{equation}
where the exponent and mantissa field of $e^{z'}_{float}$ is $u-1$ and $1+v$, respectively. Since $e^{z'}$ is always positive, the sign field is set to 0.

\subsection{Hybrid Adder Tree}


After computing the outputs $e^{z'}_{float}$ from the hybrid exponent unit, the subsequent step involves summing these values together. This summation is essential for calculating the denominator in Equation~\ref{Softmax_2}. However, it's worth noting that the summation operation among floating-point numbers can be computationally expensive. 
To reduce this computational cost, a floating point to fixed point converter (FP2FX) converts $e^{z'}_{float}$ to $e^{z'}_{fixed}$ to perform the summation in fixed-point format.
Furthermore, since the inputs $z'$ are non-positive, the values of $e^{z'}$ fall within the range of $(0,1]$. As a result, the sign bit in the fixed-point representation $e^{z'}_{fixed}$ is unnecessary, and only one bit for the integer part is required. We also implements a fractional bits of $e^{z'}_{fixed}$ to be fully configurable to enable an better flexibility between hardware performance and accuracy performance. The resultant summation $\sum_{j=0}^{N-1} e^{z'_{j}}$ is then converted back to floating point format using a leading one detector (LOD) to facilitate the operation of the hybrid division-multiplication unit (Hybrid DIV-MUL unit in Figure~\ref{fig:sm_arch}), which is described next.


\begin{figure}
\captionsetup{font=small}
    \centering
    \includegraphics[width=0.5\textwidth]{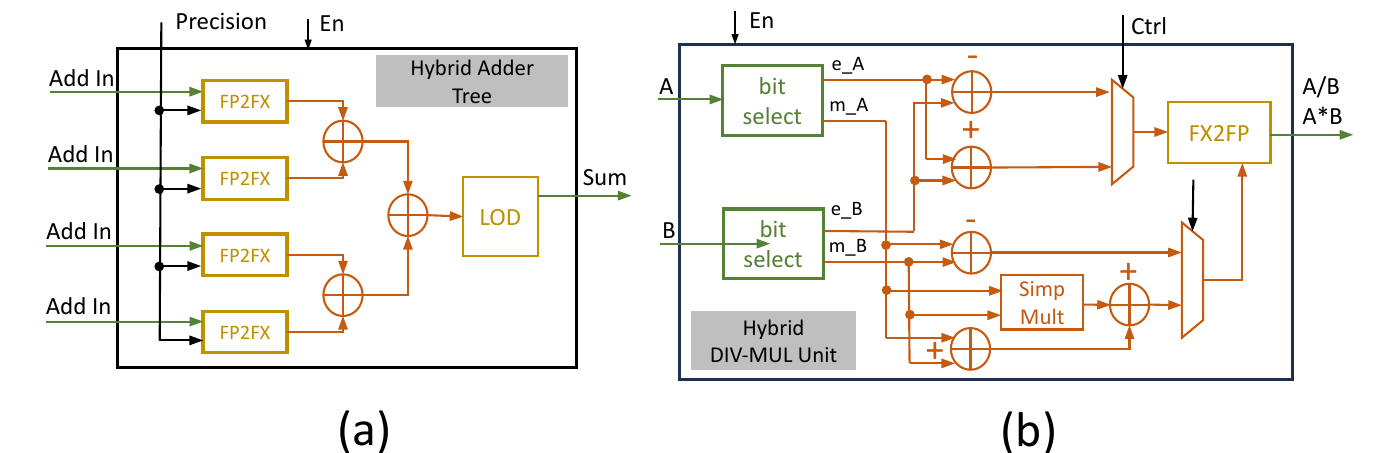}
    \caption{(a) Hybrid adder tree and (b) division-multiplication unit. }
    \label{fig:hy_exp}
        \vspace*{-2\baselineskip}
\end{figure}

\subsection{Division/Multiplication Unit} \label{sec:div-multi-unit}

Following the generation of the denominator and numerator in Equation~\ref{Softmax_2}, the next step involves computing the division between them. Division is a computationally intensive operation, and several approaches have been proposed in prior research to mitigate the associated computational cost. For instance, when simplifying the Softmax function in CNNs, the authors of \cite{aggresive_sm} and \cite{isca_sm} approximated the divisor as a power of two integer, effectively substituting the division operation with a shift operation. However, this approximation introduces errors in Softmax calculations, resulting in reduced accuracy for transformers. An alternative method by Vasyltsov et al.~\cite{klut} involved the use of a lookup table (LUT) to optimize division computation, but it relies on prior knowledge of the input distribution. In our study, we implement the log-subtract strategy as proposed in~\cite{nuaa}. However, we deviate from using fixed-point format, which can result in high computational costs for division operations. Instead, we opt for floating-point format, effectively reducing quantization errors and mitigating hardware costs. Given that both numerator and denominators are represented in floating-point format, let $e_{a}$, $m_{a}$ and $e_{b}$, $m_{b}$ represent the exponent and mantissa fields of the numerator a and b, respectively. The division $\frac{a}{b}$ can be detailed as follows:
\begin{equation} \label{div}
\begin{split}
        \frac{a}{b} & = 2^{e_a}(1+m_a)/ 2^{e_b}(1+m_b)  = 2^{e_a-e_b+log_2(1+m_a)-log_2(1+m_b)}\\
         &\approx 2^{e_a-e_b+m_a-m_b} \approx 2^{e_a-e_b}(1+m_a-m_b)\\
\end{split}
\end{equation}
where we have employed the Taylor approximation, specifically $log_2(1+x) \approx x$ for $x$ in the range of [0,1], to simplify the computation. 
While \cite{klut} demonstrated that logarithm approximation led to decreased accuracy in the transformer model, this was primarily due to the limited precision and range that an 8-bit fixed-point format can represent, resulting in the accumulation of quantization and approximation errors. In our system, the utilization of a 16-bit floating-point format can effectively address this limitation and result in negligible accuracy loss.


In the fixed-point approach~\cite{nuaa}, it requires two LODs and three shifters to convert the input to a power-of-2 format, perform subtraction, and then convert back to fixed-point format. In contrast, the floating-point system requires no shifters or LODs for Equation~\ref{div} since the input and output are already in power-of-2 format, reducing resource usage and latency.

\subsection{Softmax Backpropagation For Training}
\label{sec:backprop}
The computation of the Softmax backward propagation (Equation \ref{sm_derivative}) requires the calculation of $ss^{T}$, where $s$ represents the results obtained from forward propagation. As each element of $s$ is expressed in floating-point format, this computation involves the multiplication of two floating-point numbers as follows:
\begin{equation} \label{mul}
\begin{split}
        a \times b & = 2^{e_a}(1+m_a) \times 2^{e_b}(1+m_b) = 2^{e_a+e_b}(1+m_a+m_b+m_am_b)
\end{split}
\end{equation}

To handle this multiplication, we employ the division/multiplication unit detailed in Section~\ref{sec:div-multi-unit}. Specifically, in comparison to Equation \ref{div}, Equation \ref{mul} introduces only one additional operation, which is $m_am_b$. This operation can be efficiently processed using a fixed-point multiplier. For the remaining operations, such as exponentiation and mantissa addition, we can utilize the existing hardware within the division/multiplication unit for processing. To further save on computational resources, rather than utilizing a full-range multiplier, we opt to take advantage of only half the range of one of the multiplicands. This approach leads to a $50\%$ reduction in hardware compared to a full-range multiplier without any accuracy degradation. With this method, the division/multiplication unit can handle the computations within both forward and backward propagations, further improving the hardware utilization rate.


\subsection{Hyft Vector Processor}
\label{sec:hyft-vec-engine}

As described in the earlier sections, Softmax computation typically involves three stages of processing: (1) maximum searching, (2) exponentiation and summation, and (3) division. These three stages cannot be pipelined for a single vector operation. However, unlike traditional CNNs that execute Softmax only in the last layer, transformers require Softmax computation for multiple input vectors within an attention block. This opens up the possibility for a vector-wise pipeline, which has not been explored in previous work. Hyft is divided by the 3 stages of Softmax and each stage will process different vectors for better utilization and throughput. 

\section{Experimental Results}

\subsection{Accuracy Evaluation}
\label{sec:software_eval}

To enable efficient Softmax computation, we have incorporated several approximations, as elaborated in Section~\ref{sec:hyft-arch}. In this section, we explore the impact of these approximations on the training and inference accuracy of Hyft. Specifically, we emulate the software behavior of Hyft by implementing it in PyTorch. We then fine-tune a BERT model~\cite{devlin2019BERT} on various downstream tasks specified in the GLUE and SQuAD benchmarks. Subsequently, we replace the Softmax layer in the resulting model with the customized Softmax implementation using Hyft and record the changes in accuracy performance. Specifically, we consider two configurations of Hyft: Hyft16 and Hyft32. Hyft16 and Hyft32 accept inputs and produce outputs in FP16 and FP32 formats, respectively. We also report the BERT accuracy with original Softmax implementation.

The results are presented in Table~\ref{tab:hyft_err_tab}. We notice there is a negligible difference on the accuries with the BERT model with original Softmax implementation, indicating that both versions of Hyft can integrated well with the transformer.

\begin{table}
\captionsetup{font=small}
    \small
    \centering
    \begin{tabular}{|ccccccc|}
    \hline
            Tasks    & SQuAD 1.1      &     MRPC     &    CoLA       &  RTE       &    SST2   & QNLI \\
     \hline
         Original        &    88.32\%     &    86.03\%    &  53.31\%    &    67.02\%   &   93.24\%  &  91.10\% \\
         
         Hyft32     &    88.31\%     &    86.12\%     &   53.28\%    &    67.00\%   &   93.29\%  &  91.02\% \\
         
         Hyft16     &    88.37\%     &    85.97\%    &   53.40\%   &    66.91\%   &   93.21\%  &   91.08\% \\
         ~\cite{base2sm}   &    81.96\%   &    80.04\%    &   44.43\%   &    53.89\%   &   82.92\%  &   84.60\% \\
         ~\cite{isca_sm}     &    88.30\%     &    83.57\%    &   53.23\%   &    66.72\%   &   92.08\%  &   90.01\% \\
         \hline
    \end{tabular}
    \caption{Accuracy results of Hyft on GLUE and SQuAD benchmarks. We simulate the prior works~\cite{base2sm,isca_sm} over BERT for comparison.}
    \label{tab:hyft_err_tab}
    \vspace*{-2\baselineskip}
\end{table}

Moreover, to demonstrate the robustness of Hyft for transformer training, we assess its accuracy performance for fine-tuning the BERT model. To be specific, we substitute the original Softmax layers in BERT with our customized Softmax implementation using Hyft16 or Hyft32. We then fine-tune BERT on the GLUE and SQuAD datasets for 3 epochs, employing an initial learning rate of 5e-5 and a batch size of 32. 

The resultant training accuracies are presented in Table~\ref{tab:hyft_training}. Similarly, we observe that Hyft does not have any noticeable impact on training accuracy, showing that both Hyft16 and Hyft32 can be integrated to transformer training and inference.

\subsection{Hardware Evaluation}
\label{sec:hardware_eval}

\begin{figure*}
\captionsetup{font=small}
    \centering
    \hspace*{-1\baselineskip}    \includegraphics[width=0.87\textwidth]{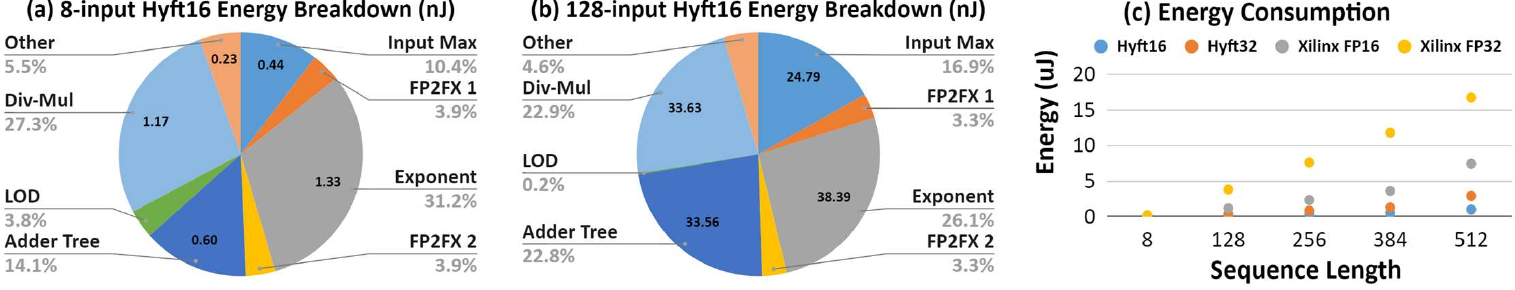}
    \caption{Energy analysis for Hyft across multiple input sequence lengths.}
    \label{fig:energy_analysis}
\vspace*{-1\baselineskip}
\end{figure*} 

We implement Hyft and evaluate the hardware performance on the Xilinx xc7z030 FPGA chip utilizing the Xilinx Vivado Design Suite.

Regarding the Softmax operation, we fix the length of the input vector $z$ to be 8 and assess all the designs based on the following criteria: 1. Processing latency of the Softmax operation. 2. FPGA resource utilization, measured in terms of Look-Up Tables (LUT) and Flip-flops (FF). 3. The maximum achievable operating frequency ($F_{max}$). 4. FOM, a metric introduced in~\cite{isca_sm}, defined as:
\begin{equation} \label{FOM}
FOM=\frac{F_{max} \times N \times W}{LUT+FF}
\end{equation}
where $W$ denotes the precision of the $z_{i}$. FOM serves as a comprehensive indicator that assesses the overall performance of Softmax accelerators. A higher FOM value signifies better performance. 

We compare Hyft with multiple baseline methods, the results are shown in Table~\ref{tab:fpga_result}. 
Among these baselines, TCAS-I'22 FP\cite{base2sm} and ISCAS'23 FP \cite{isca_sm} are two state-of-the-art Softmax accelerators. Their original works only support fixed-point numbers. To make the comparison fair, we add Xilinx fixed-point to floating-point conversion IPs and implement them in the same hardware platform as Hyft. Xilinx FP16 and Xilinx FP32 are 16-bit and 32-bit floating-point Softmax engines implemented using Xilinx IPs.

\begin{table}
\captionsetup{font=small}
    \small
    \centering
    \begin{tabular}{|ccccccc|}
    \hline
            Tasks    & SQuAD 1.1      &     MRPC     &    CoLA       &  RTE       &    SST2   & QNLI \\
     \hline
         Original   &    88.32\%     &    86.03\%    &  53.31\%    &    67.02\%   &   93.24\%  &  91.10\% \\
         
         Hyft32     &    88.26\%     &    86.06\%     &   53.42\%    &    67.08\%   &   93.13\%  &  91.19\% \\
         
         Hyft16     &    88.30\%     &    86.03\%    &   53.30\%   &    67.10\%   &   93.16\%  &   91.21\% \\
         \hline
    \end{tabular}
    \caption{Training accuracy of Hyft on GLUE and SQuAD benchmarks}
    \label{tab:hyft_training}
    \vspace*{-3\baselineskip}
\end{table}
In contrast, Hyft not only consumes $10\times$ fewer resources and operates at a $40\%$ higher frequency compared to the Xilinx FP design, but also surpasses the performance of previous fixed-point accelerators. The primary reason for this superior performance is the adaptive precision selection mechanism supported in Hyft. 

While~\cite{base2sm} achieves a high FOM, it is important to note that their Softmax implementation is tailored for CNNs and relies entirely on fixed-point numbers. This approximation is acceptable for CNNs.
Nevertheless, when integrated into transformer models, this approach can lead to notable accuracy degradation, as demonstrated in Table~\ref{tab:hyft_err_tab}. 
Although ~\cite{isca_sm} also achieves a comparable latency and FOM, ~\cite{isca_sm} adopts a very aggressive approximation for Softmax, further leading to a larger accuracy degradation than Hyft. 

To analyze the impact of input sequence length, we implement Hyft with 128 input length. Due to FPGA resource limitation, we implemented Xilinx FP16 with 64 inputs and Xilinx FP32 with 32 inputs as comparison baselines. As shown in the last four rows of Table ~\ref{tab:fpga_result}, Hyft still achieves the lowest processing latency with significantly reduced resource usage.

Figure \ref{fig:energy_analysis} (a) and (b) show the energy breakdown of Hyft16 with input sequence length of 8 and 128, respectively. We observe that the Hybrid Exponent unit and Div-Mul unit consume the most energy. Compared with 8-input Hyft, the energy consumption of input max search and exponent sum adder tree of 128-input Hyft increase significantly. While exponent and division are element-wise operations, their energy consumption percentage decreases as the input vector gets longer. 
We also compare the energy consumption of Hyft and Xilinx FP Softmax under varying input sequence length. As shown in \ref{fig:energy_analysis} (c), the energy saving of Hyft becomes more significant as the input sequence length increases. 

Based on the results of Table~\ref{tab:fpga_result}, 128-input Hyft16 can achieve $6\times$ speed up compared to Xilinx FP16. As shown in Figure~\ref{fig:softmax_runtine}, Softmax computation will contribute to $20\%$ of the end-to-end Big-Bird model runtime on an A100 GPU. With Hyft integrated to A100 GPU, it is estimated to achieve a $13\%$ end-to-end speed-up for Big-Bird model with 4096 input sequence length.

\begin{table}
    \tiny
    \captionsetup{font=small}
    \centering
\resizebox{0.48\textwidth}{!}{%
\begin{tabular}{|c c c c c c c c|}
    \hline
            Methods                     &   Config.    &   Format      &     Resource        & $F_{max}$ &   Latency    &   FOM     & Energy        \\
                                        & N, ~~ W &                    &   (LUT, FF)    &   (MHz)   &   (ns)       &           & (nJ)          \\
        \hline 
         APCCAS'18\cite{zfw2018}        &    8 16-bit   &   Fixed      &   2564, 2794     &    436    &   NA         & 10.416    &  NA           \\
         ISCAS'20\cite{nuaa}            &    1 16-bit   &   Fixed      &   2229, 224      &    154    &   NA         & 1.004     &  NA           \\
         TCAS-I'22 FP\cite{base2sm}     &    10 16-bit  &   Floating   &   2296, 2380     &    500    &   NA         &17.106     &  NA           \\
         ISCAS'23 FP\cite{isca_sm}       &    8 16-bit  &   Floating   & 1803,1264         &    476    &   16.8       & 19.866    &  6.139        \\
         Xilinx FP16                    &    8 16-bit   &   Floating   &   3146, 4281    &    500    &   44.7         &  7.181    &  43.648        \\
         Xilinx FP32                    &    8 32-bit   &   Floating   &   12269, 14198   &    454    &   89.1      &  4.391    &  171.963       \\
         \textbf{Hyft16}                &    8 16-bit   &   Floating   &   1172, 1361     & 645       &   15.5       &  32.594   &   4.278        \\
         \textbf{Hyft32}                 &    8 32-bit   &   Floating  &   2399, 2728     &    526    &   19         & 26.264    &   7.296        \\
         \hline
         Xilinx FP16                    &    64 16-bit   &   Floating   &   40489, 61304    &    500    &   80         &   5.029   &  579.952        \\
         Xilinx FP32                    &    32 32-bit   &   Floating   &   69565, 90256   &    454    &   151      &    2.909  &  950.565      \\
         \textbf{Hyft16}                &    128 16-bit   &   Floating   &   19026, 22090     & 645       &   27.9       &  32.127   &   137.291        \\
         \textbf{Hyft32}                 &   128 32-bit   &   Floating  &   46706, 57781     &    526    &   34.77         &  20.619   &   413.113       \\
         \hline
    \end{tabular}}
    \caption{Hardware evaluation of Hyft.}
    \label{tab:fpga_result}
    \vspace*{-4\baselineskip}
\end{table}

\section{Conclusion}
In this study, we introduce Hyft, a hardware-efficient floating-point Softmax accelerator designed for both training and inference purposes. Our evaluation results demonstrate that Hyft achieves outstanding outcomes, including a remarkable reduction in hardware resource utilization and processing latency while preserving transformer accuracy to a negligible extent.

\bibliographystyle{./bibliography/ACM-Reference-Format.bst}
\bibliography{./bibliography/refs.bib}

\end{document}